\pdfoutput=1

		\documentclass{isprs}
		\usepackage{hyperref}
		\usepackage{subfigure} 

		\usepackage[utf8]{inputenc}
	
		\usepackage{varioref} 
	 
		\usepackage{natbib} 
	
	\usepackage{enumitem}
		\usepackage{graphicx}
		\DeclareGraphicsExtensions{.pdf,.jpg,.png}

		\usepackage[usenames,dvipsnames]{color}
	\usepackage{amsmath}
		\usepackage{algorithm2e}
		\usepackage{microtype}  
		\usepackage{SIunits} 
		\usepackage{listings}
		\usepackage{tabularx} 
		\usepackage{setspace}

	\newcommand{\myimageHL}[4]
	{
		\begin{figure} [ht!]
			\begin{center}
				\includegraphics[width= #4 \linewidth,keepaspectratio]{#1}
				\caption{#2}  
				\label{#3}
			\end{center}
		\end{figure} 
	}

	\newcommand{\myimage}[3]
	{
		\myimageHL{#1}{#2}{#3}{1} 
	}

	\newcommand{\myimageFullPageWidth}[3]
	{
		\begin{figure*} []
			\begin{center}
				\includegraphics[width=\textwidth,keepaspectratio ]{#1}
				\caption{#2}  
				\label{#3} 
			\end{center}
		\end{figure*} 
	}

	\newcommand{\myimageURL}[4]
	{
		\begin{figure} [ht!]
			\begin{center}
				\href{#4}{\includegraphics[width=\linewidth,keepaspectratio]{#1}}
				\caption{\protect\href{#4}{#2. \protect\url{#4}}} 
				\label{#3}
			\end{center}
		\end{figure} 
	}

	\usepackage{rotating}

\title{Interactive in-base street model edit: how common GIS software and a database can serve as a custom Graphical User Interface}
\author{Rémi Cura  $^{A}$, Julien Perret $^A$, Nicolas Paparoditis  $^A$}

\address{ $^A$  Université Paris-Est, IGN, SRIG, COGIT \& MATIS, 73 avenue de Paris, 94160 Saint Mandé, France\\
	first\_name.last\_name@ign.fr
	}

\begin{document}
 



\abstract{ 
Our modern world produces an increasing quantity of data, and especially geospatial data, with advance of sensing technologies, and growing complexity and organisation of vector data.
Tools are needed to efficiently create and edit those vector geospatial data.
Procedural generation has been a tool of choice to generate strongly organised data, 
yet it may be hard to control.
Because those data may be involved to take consequence-full real life decisions, 
user interactions are required to check data and edit it.
The classical process to do so would be to build an adhoc Graphical User Interface (GUI) tool adapted for the model and method being used. This task is difficult, takes a large amount of resources, and is very specific to one model, making it hard to share and re-use.

Besides, many common generic GUI already exists to edit vector data, each having its specialities.
We propose a change of paradigm; instead of building a specific tool for one task, we use common GIS software as GUIs, and deport the specific interactions from the software to within the database.
In this paradigm, GIS software simply modify geometry and attributes of database layers, and those changes are used by the database to perform automated task.

This new paradigm has many advantages. The first one is genericity.
With in-base interaction, any GIS software can be used to perform edition, whatever the software is a Desktop sofware or a web application.
The second is concurrency and coherency. Because interaction is in-base, use of database features allows seamless multi-user work, and can guarantee that the data is in a coherent state.
Last we propose tools to facilitate multi-user edits, both during the edit phase (each user knows what areas are edited by other users), and before and after edit (planning of edit, analyse of edited areas).

}

\maketitle 
\myimageFullPageWidth{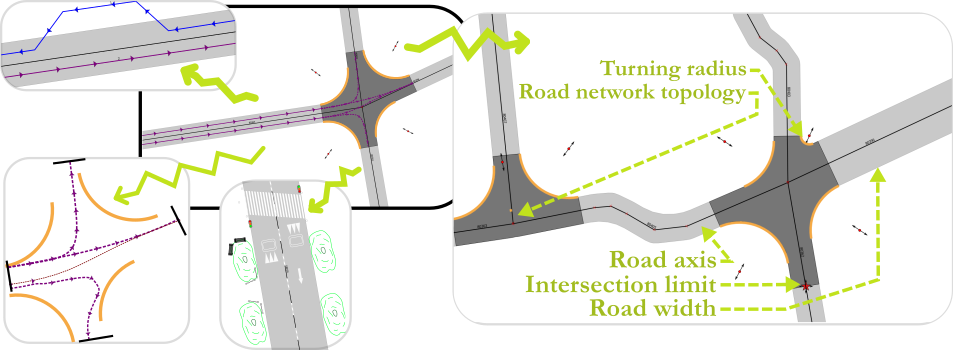}{Interaction is handled in-base rather than in custom software. Street model is regenerated automatically (StreetGen) when user edit street model parameters using convenient and effective interactors. Road model, traffic information and street features can be edited.}{usr.fig.road_edit}

%

\section{Introduction}  


Our modern world need an increasing quantity of data, and especially geospatial data.
Indeed, our capabilities to sense our environment as increased with ever more precise satellite imaging, LIDAR scanning, and mobile mapping.
In parallel, another trend tends to connect data and semantize it (semantic web), 
with more abstract data such as vector data, becoming more accessible.

The challenge we face is then to design tools to efficiently create and edit those vector geospatial data.
Generating high quality structured data is a challenge for which procedural tools are well adapted.

Procedural modelling is a powerful generative method, but notoriously hard to control (see \cite{Chen2008,Lasram2012a,Lipp2011} for examples of increasing control).
Hard control comes from the fact that understanding the link between initial parameters and the resulting model may not be obvious.
Modelling is a process of simplification, as the goal is to model a complex phenomenon with a comparatively simple model.

However, having the capabilities to model something is one thing, finding the best parameters of the model so it best fits a set observation is another.
The latter is called Inverse Procedural Modelling.
The way to find the parameters may be a sophisticated mathematical method (\cite{Martinovic2013, Musialski2013}), or a user! 
Moreover, whatever the level of automation, some user control is necessary, be it to validate and correct results, or to extend it beyond the limits of the procedural tools used.


Yet numerous non-procedural tools exist to edit geospatial data : GIS software.
Even considering only open source software, several major GIS software exist.
Unsurprisingly, each has strong points.
For instance QGIS\footnote{\url{www.qgis.org}} has a user friendly interface and can integrate a great number of other open sources tools via plugins, GRASS GIS\footnote{\url{https://grass.osgeo.org}} scales very well, can be automated and has extensive raster processing, OpenJump\footnote{\url{www.openjump.org/}} is light and has specialized tools for topology edition and validation.
Leaflet\footnote{\url{http://leafletjs.com/}} or Openlayer\footnote{\url{http://openlayers.org/}} allow to easily build custom light web clients to access and edit data through a browser.

Those tools have their specificities, and it would be pointless to try to create a super-tool grouping all others, as modern programming paradigm tend toward simplicity (KISS principle).
Users prefer to use several complementary tools to perform various tasks.
However, each one of these software applications have their own programming language, User Interface (UI), and specific way to customize it.
However they all have a common capability, which is to edit vector geometry and attributes.

We propose to take advantage of this common capability to use GIS software as interfaces for complex user interaction. 
Rather than having to create custom interaction handling for each GIS software,
we deport the interaction handling inside of the database. 

This approach might be coherent with recent trend to have lighter client that do not require installing (browser-based client).

This new paradigm can be used for many interactions, we use it to control an in-base Procedural Street generation method (StreetGen).
As the goal is interaction, speed is important, with ideal speed under $300 \milli \second$ (not noticeable), with occasional spikes of a few seconds allowed.

In this work we will use both "edition" and "digitization" as the action of editing a vector layer (both geometry and attributes).

\subsection{Plan} 
In section \ref{usr.method} we further introduce the method and the proposed in-base interaction, with details on patterns to facilitate design of in-base interaction and advanced interaction to help teamwork.
In section \ref{usr.result} we illustrate how those design patterns can be used for controlling StreetGen and facilitate edits. Section \ref{usr.discussion} introduces perspectives and limitations, and Section \ref{usr.conclusion} concludes this chapter.


\section{Method}
\label{usr.method}
In this section, we start by introducing the need to interaction and control for procedural modelling methods.
Then we introduce the in-base interaction concept, where the specific part of interaction handling is moved from the software to the database (Fig. \ref{usr.fig.in_base_interaction}).
We present basic design patterns for in-base interaction associated with examples. 
Last we consider how in-base interaction could be used to help digitization, and to help plan it and analyse it afterwards.
 
\subsection{Control of procedural modelling}
	Control of procedural generation tools have limited their use for a long time.
	Indeed, the classical workflow would be to use a procedural tool to generate a model, then manually edit the results for final details.
	Lets take the example of a drawing software. The goal is to generate a nice cloudy sky.
	Realistic clouds can be generated procedurally (using Perlin noise for instance).
	Once the user finds the proper parameters of the procedural clouds, he switches to fine editing, using brushes, erasers and so on to perfectly adjust clouds.

	However, this approach has two major issues. 
	The first is that manual edits are lost if the user wants to change the parameters of the procedural tool. It greatly reduces the re-usability, parameters exploration, sharing, etc.
	
	The second issue is more modelling specific. 
	When the user starts manual edition, the result no longer obeys the model of the procedural tool.
	This might no be an issue for drawing, but if the procedural tool generates a driving network, inconsiderate edition outside the procedural tool might break the topology of the driving graph or introduce errors.
	The obvious advantage is that by unconstraining the last human edition step, the result is not limited by the modelling space of the procedural tool.
	
	We choose another approach where the user only makes changes through the procedural tool.
	We first automatically generate a modelling ('best guess'), then let the user tunes parameters of the model,
	as well as overrides some of the automated results.
	Each time the user changes something, the relevant part of the model is re-generated at an interactive rate.

\subsection{In base interaction concept}
	\label{usr.method.interaction_concept}
	
	\myimageFullPageWidth{"illustrations/chap4/in_base_interaction/in_base_interaction"}{New proposed User Interface paradigm for GIS software. Instead of building several custom interactions for each data accessors (desktop GIS, browser GIS, etc.), we propose to use their basic vector editing (standard) and create custom interaction inside the database.}{usr.fig.in_base_interaction}
	
	We propose a new paradigm for custom user interaction in GIS software (See Fig. \ref{usr.fig.in_base_interaction}).
	Traditionally, when a custom interaction is needed, GIS softwares have to be amended, often by adding a plugin, or by coding the desired interaction (web GIS).
	Custom interactions are therefore costly and limited to one tool. Indeed, wanting the same custom interaction for several GIS software means creating the same interaction several time so it is adapted to each GIS software.
	Furthermore, each custom interaction parts have to be maintained while the GIS software evolves.
	
	For simple interaction, we propose a much simpler solution, which is to move the custom interaction handling from GIS software to database, and use the classical GIS editors (vector edition, geometry and attributes) to trigger those custom interactions.
	Thus, the custom interaction becomes available to any GIS software able to edit a vector in the database, thus nearly universal.
	
	Lets take the example where a user needs a way to create grids. The classical solution would be to create a QGIS plugin (for instance) with dedicated buttons and forms to create the grid and manage it. Such a plugin would range from simple to complex, depending on how well the grids can be managed (grid fusion, etc.). The actual QGIS functionality for grids has about 15 buttons and forms. Both the UI and actual grid creation are tailored for this GIS software.
	On the other hand, we could automate this grid creation so that modifying a standard polygon layer produces and controls the grid (See Video \ref{usr.fig.url_lens_hexagon_map}). Then, grid creation could be performed from any GIS.
	
	For simpler synchronising tasks, in-base interaction are even more powerful.
	For instance, lets consider a point layer with two orientation fields, one expressed in degrees, one in radians.
	Those fields have to be synchronised at all times. One solution would be to write custom handling in the GIS software, so that any change on one orientation is also done on the other.
	However, any changes of orientation done outside this GIS software would not synchronise orientations, thus leaving the data in an incorrect state.
	Yet programming this kind of synchronisation in-base is extremely easy and efficient,
	it also warranties that the two orientations are always going to be seen as synchronised (ACID).

	More complex in-base interactions may be needed than synchronizing two data values.
	Indeed, for inter-dependent values, special care must be taken to avoid useless computing and possibly circular references.

\subsection{Different in-base interaction types}
	\label{usr.method.interaction_types}
	In-base interaction relies on triggers: functions that get executed when a table/view is modified.
	Thus, the mean of interaction is fixed.
	However, to reach scalable and safe interactions, adapted design patterns are needed.
	In this section we introduce those basic design ideas, which are not limited to StreetGen but are generic.
	In following section,those patterns are then combined to create concrete advanced interactions for a specific application (in-base street generation with StreetGen).
	(See Fig. \ref{usr.fig.basic_design_patterns}).
	\myimageHL{"illustrations/chap4/basic_design_patterns/basic_design_patterns"}{Various design patterns for in base user interactions. In "Direct Edition", a trigger intercepts data. In "Proxy View" a view is used as a man-in-the-middle to avoid cyclic trigger call. In "Geometric Control", another geometric object is used as a control (slider, etc.) for the targeted table. Last, storing "User Inputs" in separate table and combining it with automatically generated results solves the user input persistence problem.}{usr.fig.basic_design_patterns}{1}

	\subsubsection{"Trigger in the middle"}
	The simplest form of interaction is when a user directly modifies a table content.
	Such a modification is then processed by a trigger before the modification is applied (Trigger is between user and table, hence "trigger in the middle").
	As such, it is possible to check and/or correct values modified/inserted by the user.
	
	Lets take for example the multi-users tracking system we implemented as a QGIS plugin (See Section \ref{usr.method.multiusers}).
	In this system, the position and extent of the qgis user view is registered each time the user moves on the map (screen rectangle),
	which allows to know were the user is working, and prevent persons from working on the same area without knowing it.
	We observed that user editing data with QGIS never edit objects in the corner of their screen. Indeed, they tend to move the map so the object that was in the corner is approximately in the centre of the display.
	As such, the map seen on the screen (rectangular) is not really the potential edit area, a rounded rectangle would be more appropriate.
	We create a trigger which rounds the rectangle when the rectangle is inserted into the database. See Figure \ref{usr.fig.rounded_rectangle} and web video \url{https://www.youtube.com/v/grlkUvvSf3w?hd=1&start=120&end=134&version=3}
	
	\myimageHL{"illustrations/chap4/rounded_rectangle/rounded_rectangle"}{The position of the user map extent is recorded as a potential edit area. We notice users never edit features in the corners, which means corners are not potential edit area. We then create a trigger to round the incoming rectangle in the database, so as to have more "realistic" potential editing area.}{usr.fig.rounded_rectangle}{1}.
	  
	\myimageURL{"./illustrations/chap4/rounded_rectangle/proxy_video_PAPERPRINT"}
	{Video of automatic tracking of probable editing area via QGIS and  PostGIS}
	{usr.fig.url_rounded_rectangle}
	{"https://www.youtube.com/v/grlkUvvSf3w?hd=1&start=120&end=134&version=3"}
	
	Another common example is snapping. For instance, given a linestring representing a building footprint contour, we create a point that represent this building exit door. This point should always be on the contour.
	Yet when editing it, a user could move it away from the line without noticing.
	To prevent that, a trigger in the database first projects the edited point on the line before actually saving it.
	
	Maybe the most common usage is for constraint enforcement. An user could modify an attribute which is constrained. For instance modifying the road width, a trigger enforce that the road width is positive (simply taking absolute value of user input) before saving it in the base.
	Almost all the in-base interactions we present use "Trigger in the middle".
	
	\subsubsection{Direct "geometric control"}
	"Direct Modification" imply to change one by one the object involved.
	
	In some case, it may be much more powerful to use another geometry as a controller.
	The idea is fare from new, and is well adapted to database and triggers.
	
	Lets take for instance a point cloud lens, which is defined as showing all the points within the lens geometry (See \cite{Cura2015}. In this case we control which points amoing billions are displayed with a geometric controller which is the lens geometry.
	Triggers on the lens ensure that appropriate points are displayed when change occurs.
	In addition, lens attributes can also be used to control other aspects.
	For instance a lens attribute "LOD" allows to choose which amount of points are going to be displayed. 
	Another attribute "pass" allows to choose which vehicle pass to display (in terrestrial mobile mapping, the mapping vehicle may have made several pass at the same place at different time.)
	
	\myimageHL{"illustrations/chap4/lens/lens"}{A GIS visualisation lens for point cloud, showed in QGIS. A lens (polygon) position and form controls what points are displayed (among several Billions points). In addition to lens geometry, lens attributes also controls other aspect of rendering, such as Level Of Detail (LOD) or the vehicle pass (temporal filtering).}{usr.fig.lens}{1}

	Several Direct geometric control can be used conjointly, from on or several table.
	
	Another example is the hexagonal controller discussed in \ref{usr.method.multiusers}.
	The goal is to generate and edit an hexagonal grid. Rather than adding hexagon by hexagon, we propose to use a direct "Geometric Control" (a polygon table with attributes) to control the hexagonal grid. The control table contains triggers, so that upon changes the hexagonal grid is accordingly created/updated. The control layer contains an attribute 'size' which control the size of the hexagons in the hexagonal grid.
	Such automation are easy to create and greatly simplify the control of complex objects.
	(See Video \ref{usr.fig.url_lens_hexagon_map}).
	\myimageURL{"./illustrations/chap4/hex_grid/proxy_video"}{Using a polygon proxy to control an hexagonal grid (Database triggers), showed in QGIS.}{usr.fig.url_lens_hexagon_map}{https://www.youtube.com/v/grlkUvvSf3w&hd=1&start=288&end=323&version=3&vq=hd720}

	\subsubsection{Indirect "geometric control"}
	\label{usr.method.indirect_geometric_control}
	Geometric control can be pushed one step further.
	Indeed, in both previous example, the actual geometry was directly used to control the objects (points or hexagons), that is the geometry (polygon) representing the control object was directly used.
	Yet, we can use geometric controller as graphical control element, that are abstracted from the map and whose geometry is not related to any geospatial meaning, like a slider.

	\myimageHL{"illustrations/chap4/manual_intersection_limit/manual_intersection_limit"}{An indirect geometric controller permits an easy control of the position of the intersection limit which is defined by its curvilinear abscissa. Only controller changes are used to update the abscissa.}{usr.fig.manual_intersection_limit}{1}

	StreetGen manual intersection limit gives a good example of such a design (See Fig. \ref{usr.fig.manual_intersection_limit}).
	The goal is to allow the user to be able to choose the intersection limit, which is defined by a curvilinear abscissa along the road axis.
	This abscissa could be change through a form, which lacks visual feedback and is time consuming.
	Instead, we create an indirect geometrical controller that is a point which represents the limit.
	We define triggers so that a change of the controller by the user is interpreted as a change of the abscissa, which in turns triggers the regeneration of all the impacted geometries (road surface, road intersection surface, lanes, etc.).
	In this example, the controller is indirect as the abscissa definition is not based on the controller.
	More accurately, controller changes must be interpreted before having any impact.

	\myimageHL{"illustrations/chap4/altimetry/altimetry"}{An indirect geometric controller is used to change the Z values of points of a 3D linestring within a traditional 2D GIS interface (here: QGIS). The altimetry curve is both a visualisation tool and an easy edit tool.}{usr.fig.altimetry}{1}
	In another example, the indirect geometric controller is used both for visualisation and control (See Figure \ref{usr.fig.altimetry}). 
	The aim is to allow the edition of Z values of a 3D polyline $L_{3D}$ within classical 2D GIS software.
	Editing the Z value for each node of the polyline is currently difficult as very few software allow to directly edit it. 
	Furthermore, in GIS software lines are drawn in the plane (seen from the top), which totally occults the Z values.
	We propose to edit the Z values of $L_{3D}$ through the use of an indirect geometric controller which is the altimetry profile $L_{alti}$ of the $L_{3D}$ line based on $L_{3D}$ as the origin axis.
	Conceptually, for each node $N_{3D}i$ in $L_{3D}$, we create an equivalent node $N_ai$ in $L_{alti}$ so that $N_ai$ is on the perpendicular (defined on a neighbourhood) to $N_{3D}i$ at a distance of $N_{3D}i.Z - Z_{min}$, where $ Z_{min} = \min_{N_{3D}i \in L_{3D}}{Z(N_{3D}i)}$.
	The user directly visualises Z values and slope.
	The user can edit $L_{alti}$ nodes, moving them closer or farther from $L_{3D}$. Then a trigger interprets those edits in terms of new Z values for $L_{3D}$, which is then updated, and triggers a recomputing of $L_{alti}$.
	This idea is based on the hypothesis that Z values do not vary significantly faster than X and Y values (or the altimetry curb would be very far from the initial curve.).

	\subsubsection{"Proxy view"}
	\label{usr.method.proxy_view}
	As seen in StreetGen manual intersection limit example (\ref{usr.method.indirect_geometric_control}), generated geometry can make a very useful indirect geometric controller.
	However, this introduces another issue.
	
	The controller has a trigger that launches when the controller is edited. Yet, this edition is just a mean for the user to edit the actual value (manual intersection limit curvilinear abscissa in this case). Now editing this actual value will produce in turn the re-generation of the controller, so as to have it in a coherent state with the actual value.
	Yet, this re-generation of the controller is a change of its geometry, which in turn launches the trigger, etc.
	Thus, the risk is to enter an infinite loop of self calling triggers.
	The problem may be less direct, coming in the form of cyclic trigger dependencies :
	Trigger A launches Trigger B which launches trigger C which launches trigger A, etc.
	
	In the controller case, the problem boils down to be able to differentiate between a change of the controller by a user, which must be interpreted and "translated", and an automated change (generation), which is not interpreted.
	
	We propose two specialised designs to deal with this problem. 
	
	The first design we propose is the use of a "Proxy view" (or materialised view, or table), so that the user never edits the controller directly, but rather a view of the controller.
	That way, we know that changes coming to the controller are only those from automated regeneration, and changes coming to the view only come from the user.
	An additional advantage is to clearly separate the automated generation part from the human interaction part. 
	
	We illustrate this design with an indirect geometric controller that edits an intersection turning radius through a proxy view.
	The curbstone arc centre is generated from a radius value and other geometries.
	We use it as an indirect geometric controller through a "Proxy view".
	User edits the arc centre using this view, which is then interpreted as a new radius (smallest distance between controller and relevant road surfaces).
	When the new radius is updated, another trigger re-generates arc centre and other geometries based on the new radius.
	This re-generation updates the arc centre. If not using "Proxy view", it could have triggered the interpretation, entering into an infinite cyclic trigger call
	(in fact, PostgreSQL limits the number of recursions, so it simply produces an error, and not a system crash).
	The "Proxy view" allows to separate automated changes and changes coming from user.

	\myimageHL{"illustrations/chap4/editing_radius/editing_radius"}{The curbstone arc centre is generated from the radius value and other geometries. We use it as an indirect geometric controller through a "Proxy view". The user edits the arc centre using this view, which is interpreted as a new radius. Then, another trigger re-generates arc centre and other geometries based on the new radius. No infinite cyclic trigger call occurs thanks to the proxy view which separates automated changes and user changes.}{usr.fig.editing_radius}{1}

	The second design to be able to separate automated changes from user interaction is much simpler.
	At its simplest form, it amounts to require automated changes to not only change the controller but also another column in a specific way. 
	On the opposite, a user interaction will only change the controller, and not the other column. That way we can differentiate between the automated change and the human interaction.
	We use this design for indirect geometric controller for StreetGen manual intersection limit. 
	In this case a dummy column is not needed, as the controller table also possesses the controlled value.
	Therefore, we simply check that both controller geometric position and curvilinear abscissa are coherent, knowing that any automated generation will synchronizes both.

	\subsubsection{Storing user choice}
	\label{usr.method.user_choice}
	When mixing automated results and human input, 
	Human interaction persistance is essential.
	
	In StreetGen, users can modify two things : the input data (road axis, road width, etc.)
	, and some of the generation results 
	(lane direction and trajectory, intersection trajectory, intersection limit, turning radius, etc.).
	
	We consider user inputs as overrides of default values.
	As such, we propose to store user inputs in separate tables from automated results.
	This design has practical advantages.
	Because the user input is in a separate table, it becomes easy to save it, to merge several users inputs, etc.
	The user input value is stored along with a way to identify which automated result is concerned (one or several ids may be necessary).
	
	The design is simple and can be coded in two ways using pure SQL.
	
	The first is to use EXCEPT statement.
	\lstset{language=SQL,numbersep=1cm,frame=single,stepnumber=0,frame=0,tabsize=1,basicstyle=\footnotesize, numbers=right,numberstyle=\tiny ,breaklines=true}
	\begin{lstlisting}
	SELECT id, value
	FROM user_override
	EXCEPT
	SELECT id, value
	FROM automated_results ; 
	\end{lstlisting}
	
	However such statement is not handy, as the columns of the table before and after "EXCEPT" statement must match.
	
	Another more practical solution is to join user and automated tables, then use COALESCE, which might end up to be more costly but is more adaptable.
	COALESCE(value1, value2 ..) is a SQL function returning the first non null argument.
	Using this function allows to express the condition: if a user value exists for this object, use it, else use the automated value.
	
	\lstset{language=SQL,numbersep=1cm,frame=single,stepnumber=0,frame=0,tabsize=1,basicstyle=\footnotesize, numbers=right,numberstyle=\tiny ,breaklines=true}
	\begin{lstlisting}
	SELECT id,
		COALESCE( user_override.value, automated_results.value) 
	FROM automated_results
		LEFT OUTER JOIN user_override USING (id);
	\end{lstlisting}

	We illustrate user input persistance for StreetGen lane edition.
	By default, lanes are generated according to the street axis direction and their position regarding the road.
	Lane geometry is obtained by offsetting the road axis curve.
	Users can override the lane direction. In this case a new row is inserted into the user input lane table. This row stores the new value chosen by the user (the lane should be in the other direction). If the road axis is edited, a new lane geometry will still be automatically generated, as the user only override direction and not geometry.
	If the user also edits the lane geometry to customize it, the customized geometry is also stored in the user input lane table.
	Now, whenever the lane must be regenerated, its geometry will be overridden by the corresponding user input.
	If the user deletes the lane, a triggers interprets that as a command to return to the default behaviour, and the corresponding row is thus deleted in the user input lane table.

	\myimageHL{"illustrations/chap4/editing_lane/editing_lane"}{When the user overrides a default behaviour, the parameter is stored in a user input table. Objects overriden do not use default generation anymore thanks to COALESCE. The user can still delete the overriden object to return to the generic behaviour.}{usr.fig.editing_lane}{1}

\subsection{Efficient Multi-user data edit}
	\label{usr.method.multiusers}
	The work presented in this section has been achieved together with Lionel Atty (SIDT, IGN, all the python development).
	A proof of concept open source QGIS plugin is available \footnote{\url{http://remi-c.github.io/interactive_map_tracking/}}.
	
	\subsubsection{Collaborative data and Gamification}
	We observe two major trends in the last five years regarding Geographic Information edition.
	
	\paragraph{Collaborative editing}
	The first trend is toward collaborative editing.
	The success of projects like OpenStreetMap\footnote{\url{www.openstreetmap.org}} 
	have put into light the advantages of collaborative editing.
	Working simultaneously to edit data greatly increase the scaling possibility.
	Moreover, supporting several users edit may also improve quality,
	as a mix of advanced and regular users is possible,
	some user possibly having a quality control role.
	
	\paragraph{serious game}
	The other trend is more pervasive, and concerns gamification (or serious game). 
	Gamification is the integration of game-related elements into a non-game context.
	For instance earning virtual points, badges, achievement, etc.
	It has proved to be a powerful incentive that can diminish the subjective effort associated with a task.
	(See \cite{Djaouti2011} for a classification of serious game).
		
	We conceptually build on both trends to create a tool helping multi user in-base interaction.

	In order to achieve multi user editing capabilities, 
	we need first to use mechanisms so that simultaneous edits of the same data is dealt with by the database (See Chapter 3).
	That is a safeguard against computer errors.
	The ability to not crash when several users edit the same data is not sufficient for efficient multi-user editing. 
	Users also need a way to be aware of each others, and most notably of who is working (or has worked)
	on which area, which is a safeguard against human error.
	
	The first problem was introduced and solved in previous work. 
	In this section, we propose a solution for the second problem.
	The proposed methods have been implemented as a QGIS plugin with Lionel Atty.

	\subsubsection{Better server interaction with auto save and refresh}
		We introduced in-base interaction concept in \ref{usr.method.interaction_concept}.
		Advantages are numerous, including the possibility to use many clients because all the interaction happens inside the database.
		This interaction concept is based on the fact that when a user edit data or a controller, 
		it will triggers in-base behaviours.
		
		For instance, a user moves a road axis, which triggers the regeneration of associated road surface and intersections, as well as lanes and intersection trajectories.
		For this interaction to be an efficient human interface, the in-base behaviour should be clear and fast enough to be interactive (less than a second),
		and most importantly, the user should have a feedback when he performs an action.
		Based on the proposed architecture, the feedback can only happen when the database received changes on data or controller.
		Yet, some GIS software like QGIS do not send changes to database unless user specifically asks for it (via a button: saving current change inside a layer).
		This mechanism is intended to provide a local edition (including revert capabilities) before actually sending the data to database.
		Based on this, user can not receive any feedback until the changes are send to database.
		Furthermore, QGIS has no way of knowing that editing data or a controller has changed other layers,
		whose rendering should also be updated.
		
		As a proof of concept, we create a QGIS plugin to disable the local edition, so that any change on a PostGIS layer is directly sent to the database,
		 and forces refresh of all rendered objects when a change has been sent to database, with a slight delay.
		That way, all the database trigger-based interactions appear to be interactive.

	\subsubsection{Easier collaborative editing with user map tracking} 
		
		Having the database model and interaction being able to deal with more than one user is just one of the necessary steps for efficient multi user work.
		Users are human, so team work requires adequate processes and tools.
		We identified one minimal requirement to enable efficient teamwork, 
		which is to be able, at all time, to quickly see what and where other people are working on the map.
		
		We propose an approach inspired by Google Doc\footnote{\url{https://docs.google.com}} collaborative editor, 
		where the editing cursor of each user is highlighted in one dedicated colour for the other users to see.
		The idea is similarly to display the current and former area of editing of each member of a team.
		
		Each time an user (with this feature activated) browses the map between $min_{scale}$ and $max_{scale}$, this user screen extend is recorded
		in a common PostGIS layer.
		Along the screen map extend are also recorder a unique user id (session name + IP) and time (in \milli\second ) .
		A simple layer style with a random colour per user id allows then to see where each user is working (See Figure \ref{usr.fig.users_tracking}).
		
		\myimageHL{"illustrations/chap4/interactive_map_tracking/users_tracking"}{Example of multi-user screen extent tracking. Successive screen extents are recorded through time (oval geometries), a long with a user-id and precise time. Displaying those screen extents immediately informs about who works where. Potential conflicts (one user editing the same area twice or 2 users editing the same area at roughly the same time) are automatically detected and a label appears on the screen.}{usr.fig.users_tracking}{1}
		
		Those screen extents are recorded asynchronously via a stack (LIFO), so as to never slow the editing or reduce interactiveness.
		
		All users record their screen extent in the same PostGIS layer, with an unique identifier and precise time.
		This allows to create PostGIS views to warn when potential work conflicts occurs.
		We created two examples of such conflicts. 
		The first is when a user comes back on an area he edited more than 5 minutes before (potential risk of re-editing the same area twice).
		The second is when two user are editing the same place at roughly (less than 5 minutes) the same time, again potentially risking duplicate work.
		We stress that all this is only informational (no coercive ability), users still have full control.

	\subsubsection{Collaborative planing and gamification}
		
		We presented a tool to allow users where other users are working, that is to facilitate teamwork during edit time.
		However, in a real life work-flow, some planning occurs before a team edits data for a given area, and some analysis may be performed after the edit is finished.
		
		We propose to use an hexagonal to-do grid to help planning and analysis, as well as introduce a small amount of gamification.
		Before edit starts, a working area is defined (by several polygons).
		An in-base interaction (See Section \ref{usr.method.multiusers} ) generates an hexagonal grid covering the defined area.
		Each hexagon also holds information 'todo' or 'done', 'todo' by default. 
		Hexagonal tiles are red when 'todo', and blue when 'done'
		When a user map extend is saved, all the hexagons covered by this extend are set to 'done'.
		The hexagonal map is then a fun way to see what has been done and what remains to do,
		as tiles change colour.
		
		When edition is finished, the same hexagonal grid can be used as a support to display information,
		for instance the cumulated edit time.
		  
		\myimageHL{"illustrations/chap4/interactive_map_tracking/before_after_edit"}{First a working area is defined, which automatically generates an hexagonal 'todo' grid. Then when user work on an area the corresponding hexagons are marked as 'done'. Afterward, the hexagonal grid can be used ot display editing time spent per area.}{usr.fig.before_after_edit} {1}
		 
	

 \section{ Result }
	 \label{usr.result}
\begin{figure} [h!] 
	\begin{center}
	\begin{tabular}{cc} 
			\href{https://youtu.be/rBWZs50wVHg}{\includegraphics[width=0.4 \linewidth,keepaspectratio]{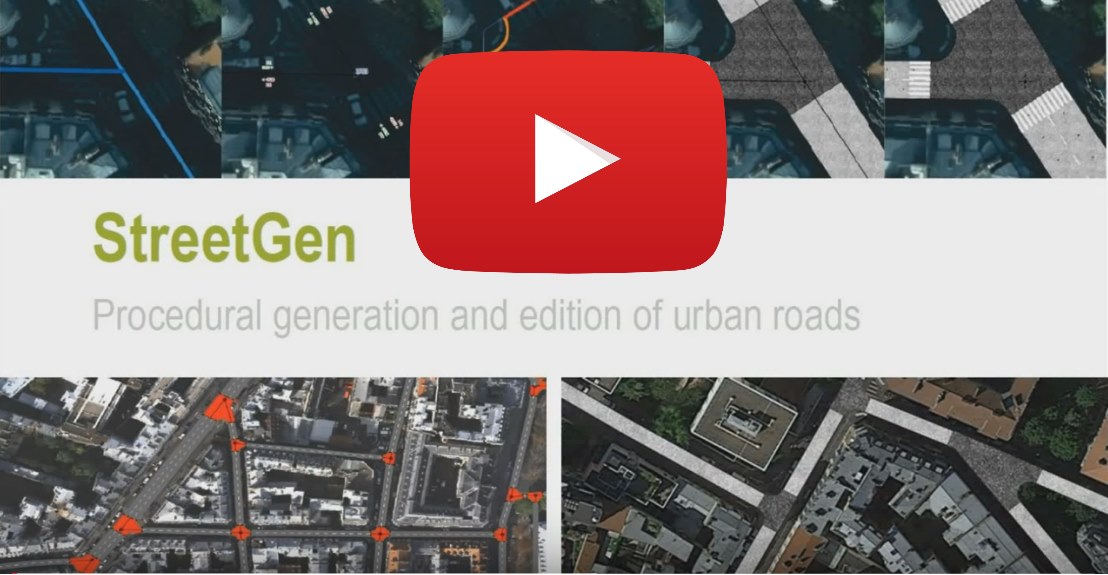}} 
			&
			\href{https://youtu.be/yIG_5MBODfo}{\includegraphics[width=0.4 \linewidth,keepaspectratio]{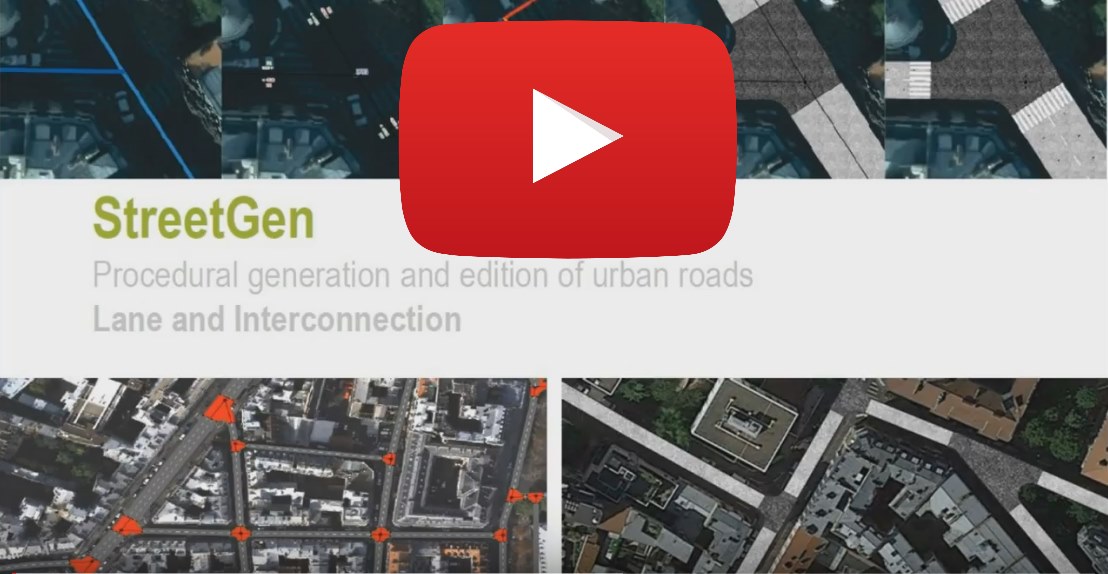}} 
\end{tabular}
	\caption{ Videos of StreetGen in-base interaction for basic parameters \url{https://youtu.be/rBWZs50wVHg} and lanes and interconnections	\url{https://youtu.be/yIG_5MBODfo} .}  
	\label{usr.result.video_link}
	\end{center}
\end{figure}

	In this section we introduce actual in-base interactions that combine 
	previously introduced patterns (see Section \ref{usr.method.interaction_types}).
	Those examples are partially extracted from StreetGen (\cite{Cura2015a}),
	an in-base tool to generate streets.
	StreetGen models several things that can be edited in different ways.
	(Videos demonstrations are available for StreetGen (See Fig. \ref{usr.result.video_link}).)
	\begin{itemize}[noitemsep,topsep=0pt,parsep=0pt,partopsep=0pt]
	\item The road information (\S \ref{usr.result.road}), which separates constant width sections and intersections, and model road surface and intersection surface based on curb stone with specific turning radius.
	\item The traffic information (\S \ref{usr.result.traffic}), with lanes and lane interconnection.
	\item The street objects (\S \ref{usr.result.street_objects}), which are semantic objects that may be defined relatively to a road axis.
	\end{itemize}
	
	\subsection{In base interaction}
		We tested the in-base interaction concept with several common open source GIS softwares
		(several versions of QGIS, OpenJump, GrassGIS ).
		In all cases, edition correctly triggers in base interaction.
		
	\subsection{Interactive road}
	\label{usr.result.road} 
	StreetGen road model parameters (See fig. \ref{usr.fig.only_road_edit}) are the road axis network topology, road axis geometry, road width, curbstone radius (turning radius) and manual override of intersection limit.
	
	\myimage{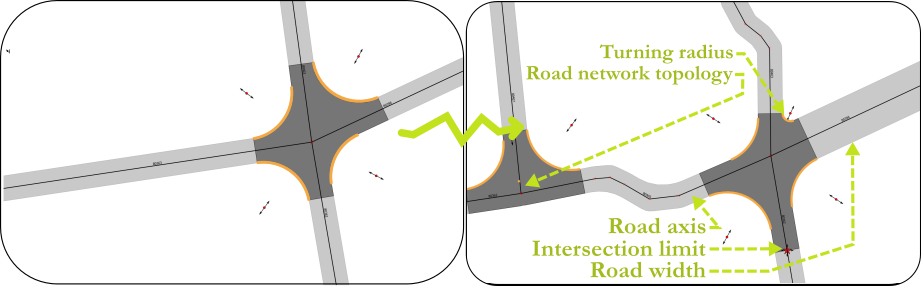}{StreetGen road model parameters, each can me modified using in-base interaction.}{usr.fig.only_road_edit}
	
	We present in-base interactions divided in two parts. The first part is the core edition, which allows to edit all parameters. The second part is improvement of core edition to create a better user interface.

	\subsubsection{Road editing}
	
		\paragraph{Editing postgis topology network}
	The very basis of StreetGen modeling is a road axis network that uses PostGIS Topology.
	Interactive topology edition is complicated, especially if topology is semantized.
	The problem stems from the necessary interpretation of user action to transcribe it into topologically valid operations.
	We implement this in-base interaction using the "Proxy view" design, 
	so as to provide a safe and dedicated interface.
	We add two views : 'edit-node' and 'edit-edge'.
	We propose a proof of concept free and open source 
	\footnote{\url{https://github.com/Remi-C/postgis_topology_edit}}.
	\\
	
	Lets take an example where the topology only contains two nodes and one edge (a line) between the nodes.
	The user creates a new node that is close to the middle of the line, but not exactly on the line. 
	This behaviour has to be interpreted has "I clicked close to the line, but in fact I meant on the line", thus this node has to be snapped to the line and the edge split into two parts, with relevant topological information updated, as well as semantic'.
	
	In more complex edition cases the expected behaviour might not be so well defined.
	Therefore, when we create in-base interactions to edit a postgis topology,
	we purposely limit the possible user actions with explicit error messages.
	We limit the interaction so that in any case we can use the postgis topology API safely.
	The main limitation is that except in obvious case, no edge split automatically occurs.
	See Fig. \ref{usr.fig.edit_topo} for example of user interaction.
	
	\myimage{"illustrations/chap4/edit_topo/edit_topology"}{Example behaviour of in base interactive topology edit, through two views, 'edit-node' and 'edit-edge'. Ambiguous user inputs are avoided.}{usr.fig.edit_topo}
	
		\paragraph{Editing StreetGen road axis network}
	
	Postgis topology interactive editing only changes topology.
	Other triggers are necessary to adapt it to the StreetGen data model, and ultimately regenerate axis that have been created/updated.
	In particular, deleted or created edges will have an impact on all of the StreetGen data model tables. This changes are propagated with a mix of trigger and using postgres foreign key (for delete).

		\paragraph{Editing of road axis}
	Road axis can be edited both for the geometry and for attributes, such as road width and number of lane.
	We use the proxy view to separate user input from automated modifications.
	In fact, any change of input data like this simply triggers a relaunch of StreetGen on the concerned elements, which unify processing.
	
	User can edit radius by typing a new value, or use the indirect "geometric controller" as seen in Section \ref{usr.method.proxy_view}.
	
		\paragraph{Editing of intersection limit}
	As seen in \ref{usr.method.indirect_geometric_control}, intersection limit uses a combination of indirect "geometric controller", "proxy view", and "sotring user choice".
	
		\paragraph{Advanced road width editing}
	Default way to edit a road width is to change the 'width' value of the road axis attribute, which is a lot of clicks and a waste of time, as most likely the user does not know the correct value, and will have to try several widths.
	
	Instead of that, we propose to use an indirect "geometric controller" to streamline road width edition.
	The idea is to indirectly provide road width by indicating where the sidewalk is. Road width is then automatically extracted from the potential sidewalk position by first assigning each border points to a road section using section surface (which can be done efficiently with geospatial indexes), then compute new width by taking the median value to the distance to the relevant road axis. Road width is then updated with the new value, and the relevant road is re-generated.
	
	\myimageHL{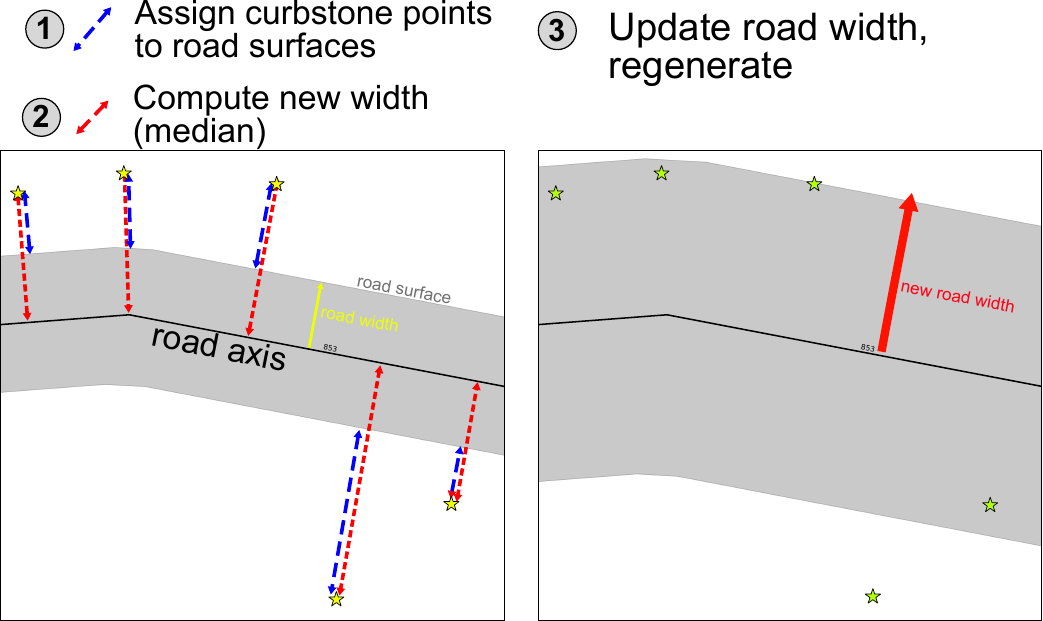}{Road width editing is extremely facilitated by the use of an indirect geometric controller. The user simply add points on the cornerstone and the road width is automatically updated, instead of guessing a width and typing it.}{usr.fig.edit_width}{1}
	
	We present in-base interactions divided in two parts. The first part is the core edition, which allows to edit all parameters. The second part is improvement of core edition to create a better user interface.
	
	In an informal experiment, we try to find the correct width of a street with and without the indirect controller. Without controller, it takes a dozen tries to find the correct width with less than 0.1 metre error (30\second). With controller, it's just one click (2 seconds).
	The advantage of the controller interface is even more obvious when road axis geometries are adjusted afterward. Indeed, if road axis is not well centred, the road width must be adjusted, as opposite to the controller version, 
	where the road width is automatically recomputed when the road axis geometry is changed. 

	\subsection{Interactive traffic}
	\label{usr.result.traffic}
	StreetGen also generates basic traffic information, such as lane geometry and direction, and interconnections between lanes in intersections.
	Both lane and interconnection use a combination of "Proxy view" and "user choice". Furthermore, interconnection also uses an indirect "geometric controller" as control point of the Bezier curve.
	
\subsubsection{lane editing}
	As seen in Section \ref{usr.method.user_choice}, the user can edit each lane direction and geometry. 
	When the user edits direction or geometry, he steps outside of automated generation (Automated results will be overridden by user edits).
	However, the user can return to automated generation by deleting the lane, which is then interpreted as a delete on user override rather than a real delete of the lane.
	Lane number is also editable through another "Proxy view" on road axis.
	 
\subsubsection{interconnection editing} 
	Interconnection have a more complex behaviour.
	In lane case, a lane always exists, because the number of lanes is a parameter.
	For interconnection, the user needs a way to convey the information that an interconnection between two lanes might not be authorized.
	For instance, at a given intersection it might physically possible to turn left, but it is forbidden by law.
	
	This information is stored in a boolean. The user does not set it directly, but instead deletes the interconnection geometry. 
	If interconnection was not overridden, then the interconnection is marked as not allowed. Else, user parameters are deleted as for the lane case.
	
	A particularity of interconnection is the use of Bezier curves to model trajectories (See Fig. \ref{usr.fig.edit_interconnection} ). 
	
	This curve is controlled by classical control points that are stored in the same table as the interconnection trajectory (default control points are implicit).
	
	\myimageHL{"illustrations/chap4/edit_interconnection/edit_interconnection"}{User editing interconnection, both for customized trajectory with bezier curve and possibility to use this trajectory.}{usr.fig.edit_interconnection}{1}

	\subsection{Interactive Street Objects}
	\label{usr.result.street_objects}
	\subsubsection{Generic street objects}
	
	Generic street objects are semantic points that can be positioned relatively to a street axis. If this is the case, object position is defined by an curvilinear abscissa and distance to street axis (or to cornerstone).
	We stress that relatively positioned object must still have a synchronised absolute position so as to be correctly displayed in GIS.
	
	Object orientation can be similarly absolute or relative to a street axis.
	When a street axis is affected (change of geometry, of width, or change of topology), all the relevant street objects are updated so as to have coherent relative and absolute positioning (and orientation).
	Object positioning types may be overridden.
	For instance, if an object is positioned relatively to a street axis,
	and this street axis is deleted (topology change), this object must be
	switched to absolute positioning.
	
	\myimage{"illustrations/chap4/objects_updating/objects_updating"}{Street objects can be defined relatively to streets. In this case, a change on street automatically triggers the re-computation of the absolute object position and orientation.}{usr.fig.street_objects_updating}
	
	This mechanism warranties that objects are always coherent.
	We also use "Proxy view" so that user can interactively create/edit/delete street object.
	At object creation, user makes a choice between relative and absolute positioning (and orientation).
	In case of relative positioning, the reference can be street axis or side-walk.
	
	Interaction handling is very complex when object are in a relative position.
	The first level of complexity comes from the necessity to synchronise relative and absolute positioning, knowing that the user can change both, and that those changes must always be transcribed into relative positioning.
	For example, an object is defined relatively to a street axis. The user moves the object in a GIS software (thus changing the object absolute position). Then trigger interprets this move as a need to update relative positioning based on new absolute positioning. Then a new absolute positioning is generated based on new relative positioning.
	In this, street object becomes its own "geometric controller ".
	
	The second level of complexity comes from implicit reference handling.
	Indeed user choosing relative positioning never explicitly indicates to which street axis the object refers.
	Instead, this axis is automatically found (closest one at creation) and updated (if an axis is split for instance, or if user moves the object very far from the street).
	For instance if the user moves an object from one street to another, the relative positioning gets updated and references the new street axis.
	
	\myimageHL{"illustrations/chap4/edit_objects/edit_objects"}{Street objects edition can be done through the change of attributes (here, with QGIS). Simply moving the object also automatically updates relative positioning information.}{usr.fig.edit_objects}{1}

	\subsubsection{Specialized street objects}
	We illustrated the possibility of specialised objects with a proof of concept example about pedestrian crossing.
	
	Specialised objects add a layer of automation on top of regular street objects.
	For instance, pedestrian crossing creation is done via a 'proxy view' strategy.
	User create a polygon roughly representing the pedestrian crossing (possibly using only 3 points).
	This polygon is then analysed to extract pedestrian crossing parameters (width and orientation).
	We could not find in the literature a method to robustly (regarding the number of points and points repartition) fit a parallelogramoid (both side of the road may be polylines).
	Therefore, we propose a simple one: each segment of the polygon envelop votes for an orientation (weighted by segment length). Final orientation is the average of the votes.
	The pedestrian crossing width is determined by separating segment into left and right of the road axis. Each side determines a width, the final width is the average width of both sides.
	
	In fact, finding the best pedestrian crossing model adapted to user inputs is already inverse procedural modelling.
	
	\myimageHL{"illustrations/chap4/edit_pedestrian/edit_pedestrian"}{ Specialised object pedestrian crossing creation is greatly facilitated by automatic parameter extraction and generation.}{usr.fig.edit_pedestrian}{1}
	
	When a user modifies a pedestrian crossing geometry, its parameters are recomputed.
	This way, one graphic controller allows to control all pedestrian crossing parameters (position, orientation, width).
	Again, this allows for easy parameter changes via "graphic control".
	
	\subsection{Efficient Multi-user data edit}
	\subsubsection{Better server interaction with auto save and refresh}
	The auto save and refresh plugin is not necessary per se, provided the frequent use convenient save and refresh short cuts.
	However it adds a great deal of comfort and reduces the number of clicks.



\section{Discussion}
	\label{usr.discussion}
	In this section we discuss elements of method (\ref{usr.method}) and result (\ref{usr.result}) sections.
	We start by analysing need of interaction for procedural modelling, and proposed design patterns.
	We examine then how the proposed method to facilitate multi-user work perform.
	
	The next part of discussion is dedicated to interaction in StreetGen, with interactive road, traffic and objects modelling.
	
\subsection{In base interaction for procedural modelling}
	We stress that the results that can be obtained by our method (interactive procedural modelling) are limited 
	by the modelling capabilities of the procedural tool.
	Our simple road model (fixed width + intersection) can not model all existing roads.
	Prominently, some road have varying width.
	Similarly, our model can not generate all type of cornerstones, for instance cornerstones using two successive radius, or chamfered.
	
	We propose to amove the interaction from client to database. 
	While it brings numerous advantages, it is ill suited for very complex interactions,
	where dedicated Human Machine Interface (HMI) would be more appropriate.
	Indeed, in base interaction can happen only \emph{after} an edit occurred,
	which prevents any HMI scenario where the HMI proposes solutions \emph{before} edit is done.
	Yet those interaction fall in the significant Guided Design trend.
	When the complexity of interaction increases, the current PostgreSQL trigger framework also becomes a serious limitation.
	Most popular User Interfaces (UI) are based on signals (for instance QT\footnote{\url{www.qt.io/}}),
	which can be rudimentary mimicked by PostgreSQL triggers.
	However, triggers offer almost no modern control, and, as such the difference with modern UI is similar to the difference between assembly languages and modern object oriented languages.
	Therefore, in base interaction scales badly in terms of code complexity, generality and maintenance.
	As such, in base interaction should be limited to straightforward cases, and not be pushed too far.

\subsection{Different in-base interaction types}
	We propose several patterns to facilitate in-base interaction,
	yet, the distinction between patterns is quite artificial, and real use cases tend to blend all patterns.
	
	Using controller and/or proxy view necessarily increase the database server workload.
	For a "Proxy view" strategy, the choice between a "VIEW", a "TABLE", or even a "MATERIALIZED VIEW" may vary a lot depending on the load, quantity of data, complexity of code maintenance, etc.
	In this article we only explore triggers for in base interaction.
	Yet, databases also have powerful rule systems that could be used for basic interaction.
	
	As a perspective, storing user choices is a first step, but more advanced features could be attained,
	like storing user choices and archiving it, so as to have access to former user choices, rather than delete/overwrite.

\subsection{Efficient Multi-user data edit}
	An obvious limitation of the auto save and refresh is to disable local undo/redo control.
	It also breaks the concept of in-base interaction as it creates client-side code. 
	The user map tracking is fuzzy by nature, as the screen map extent is registered each time the user changes it,
	regardless if an edition occurred or not.
	This can not be avoided, as sometime quality control (mostly not editing things, but checking parts of the map) is as important as edit, and should also be tracked.
	Indeed, two users performing a check on data could easily check the same area without being aware if not using the plugin.
	The gamification concept could be pushed much further, with virtual points, awards, etc.
	More importantly, a real edit work flow would benefit from more advanced tools with user having multiple roles (editor, checker, manager, etc.).
	All the role interactions could be helped by plugin, and happen in base with the hexagonal grid support.
	For instance, a team leader could assign different areas to be reviewed to his colleagues.
	After delivery, a client finding a problem could mark the relevant hexagons, so the that edit team has easy and immediate notification of erroneous area.
	We stress that although not limited in theory, we never tested the plugin with more than 3-4 users.

\subsection{Interactive road} 
	Road editing is seriously limited by the topological road axis network editing.
	Indeed, our interactive topology editing may lead to incoherent in the implicit faces of the topology. 
	This is an implementation limitation rather than a conceptual limitation.
	Currently there is no way to split many edges at the same time, to introduce a road axis cutting Paris in half for instance.
	We noticed however that this interactive topology edition is very useful compared to the alternative,
	which is to recompute the whole topology from scratch for each change. PostGis Topology is not fast, building topology for paris street is several minutes.
	In some case, we would benefit from higher level operation, like replace several small intersection with one roundabout for instance.
	
	We also noticed that introducing geometric controller for turning radius and road width is extremely helpful, with speed gains about one order of magnitude,
	and edit much more agreeable. 
	
\subsection{Interactive traffic} 
	Users can edit lane and interconnections.
	Lane direction editing could be more effective, maybe using a geometric controller.
	Users have to edit a field 'direction', which is not handy, especially when several lanes could be edited at once.
	When users delete interconnections, it actually sets the interconnection trajectory as forbidden. 
	This behaviour greatly speeds editing, because several interconnections can be selected and deleted at once.
	However, by default all possible interconnection are authorized, which creates a great number of interconnections in intersections with many lanes.
	Interaction would be much more efficient if we could use some heuristics that would connect lanes more conservatively.
	Bezier curves are great for ease of control, but rather inaccurate when it comes to actual vehicle trajectory.
	
\subsection{Interactive Street Objects} 
	
	The street object interface is especially useful as it allows to somehow compensate limitation of road model.
	For instance, StreetGen road model does not consider parking spaces, which is a strong limitation in Paris where parking spaces are omnipresent,
	and especially meaningful for urbanism.
	Yet those parking spaces could be modelled as street objects, using an adhoc object specialisation similarly to pedestrian crossing.
	We presented a specialisation for an object which is a surface by nature (pedestrian crossing), yet many street objects are also linear by nature (like some markings and barrier).
	Street objects were only tested at street scale. 
	
	Moreover, we only presented interactive editing of street objects, and not large scale generation,
	with advanced patterns and rules. Good examples of adequate complexity can be found in shape grammars designed for city generation. 
	
	\myimageHL{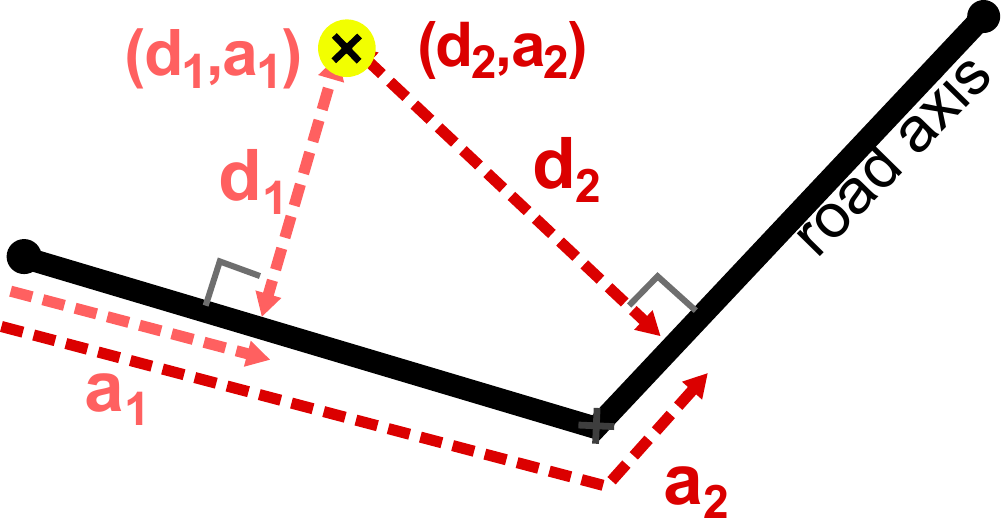}{A unique absolute position can correspond to several relative positioning, which theoretically limits relative positioning setting through absolute geometry.}{usr.fig.relative_to_absolute}{0.80}
	
	There is a very fundamental limitation to switch between absolute and relative positioning like we do (See Fig. \ref{usr.fig.relative_to_absolute}).
	Basically, going from relative to absolute positioning is not a bijection, so, in some cases there is no inverse function (going back).
	So, in some cases a couple (curvilinear abscissa, distance to axis) may not be settable through geometric proxy.
	This problem also affects the altimetry example we gave (See Fig. \ref{usr.fig.altimetry}).
	However, when special cases require it, it is still possible to set the relative positioning manually.
	
\subsection{Best of 2D and 3D world for edition}
	\label{usr.discussion.itowns_in_qgis}
	The proposed method is based on common GIS softwares, which represent and deal with data in 2D.
	The 2D view (map view) has obvious advantages for edition: it is simple, clear, and edition can be efficiently performed with usual interactive devices (mouse).
	Yet, the 2D view is sometimes confusing, especially for objects like street signs that might become invisible in 2D.
	Our brain is also extremely good at understanding 3D scenes.
	In this optic, we could propose a mixed 2D-3D edition to get the better of both worlds.
	
	We explored this idea with a prototype (work performed by Lionel Atty),
	where we use QGIS as he main 2D edition software.
	We create a plugin containing a web browser, then display ITowns\footnote{\url{github.com/iTowns/}}, a WebGL application able to show streetview, street Lidar, and perfom measures and edition.
	Both are synchronized, so that edits in 3D are also displayed in 2D, and so that 3D camera position and orientation is also controlled and displayed in 2D.
	The 3D streetview gives exemplary context awareness, and can be used to perform precise edition, while the 2D view gives good overview and fast navigation, and can be used for classic edit.
	
	\myimageHL{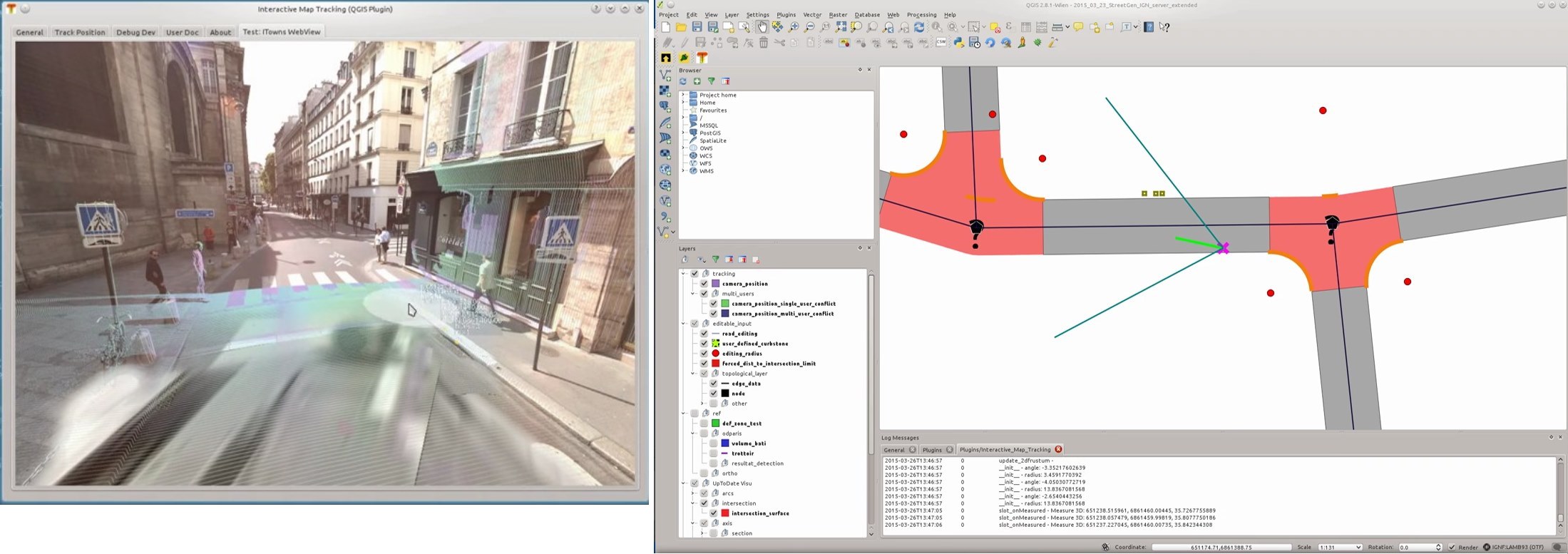}{Coupled 3D (left, ITowns) and 2D (right, QGIS) visualisation and edit, providing clear and fast edit (2D) with advanced view and 3D capabilities (3D).}{concl.fig.itowns_in_qgis}{1}



\section{Conclusion} 
\label{usr.conclusion}
In this chapter we proposed a new paradigm for custom user interaction with GIS software, where interaction handling is moved from GIS softwares to the database.
In this paradigm, GIS software simply modify geometries and attributes of database layers, and those changes are used by the database to perform automated tasks.
In the most basic form, this automated interaction can be used to check changes, for instance rounding coordinates.
The database can intercept the change and adapt it, for instance to automatically simplify a polygon.
For more complex interaction, we demonstrated the use of geometric controller, which are conceptually close to UI controller such as sliders, but are made of geometries with attributes.
Such geometric controllers can exempt a user from using a form, thus being an order of magnitude faster.
We demonstrated those capabilities with several examples of various complexity, including the interactive editing capabilities of StreetGen, an in base procedural street generator tool.

Last in base interaction can also be taken one step further and be leveraged to help team work.
In particular, work planing is possible before editing, work analysis is possible after the edition is completed (quality), and the edition can even be enhanced by introducing gamification elements.

\section{Acknowledgment}
This article is an extract of \cite{Cura2016thesis} (chap. 4). We thank Prof.Peter Van Oosterom and Prof.Christian Heipke for their extensive review.


	\section{Bibliography}  
	\bibliography{./all_bibli} 

%
%
%
%

\end{document}